\documentclass[letterpaper,nolinenumber]{jdsart}

\firstpage{538}
\lastpage{556}
\pubmonth{July}
\pubyear{2023}
\volume{21}
\issue{3}
\doi{10.6339/23-JDS1110}

\usepackage{color}
\usepackage{fancyvrb}

\DefineVerbatimEnvironment{Highlighting}{Verbatim}{commandchars=\\\{\}}
\usepackage{framed}
\definecolor{shadecolor}{RGB}{248,248,248}
\newenvironment{Shaded}{\begin{snugshade}}{\end{snugshade}}

\newcommand{\ControlFlowTok}[1]{\textcolor[rgb]{0.13,0.29,0.53}{\textbf{#1}}}
\newcommand{\DataTypeTok}[1]{\textcolor[rgb]{0.13,0.29,0.53}{#1}}
\newcommand{\DecValTok}[1]{\textcolor[rgb]{0.00,0.00,0.81}{#1}}

\newcommand{\FloatTok}[1]{\textcolor[rgb]{0.00,0.00,0.81}{#1}}

\newcommand{\KeywordTok}[1]{\textcolor[rgb]{0.13,0.29,0.53}{\textbf{#1}}}
\newcommand{\NormalTok}[1]{#1}
\newcommand{\OperatorTok}[1]{\textcolor[rgb]{0.81,0.36,0.00}{\textbf{#1}}}
\newcommand{\OtherTok}[1]{\textcolor[rgb]{0.56,0.35,0.01}{#1}}

\newcommand{\StringTok}[1]{\textcolor[rgb]{0.31,0.60,0.02}{#1}}

\usepackage[utf8]{inputenc}

\usepackage{amsmath, bm, amssymb} \usepackage{tabularx} \usepackage{booktabs} \usepackage[linesnumbered,ruled,vlined,procnumbered]{algorithm2e} \usepackage{tikz-qtree}

\tikzset{every tree node/.style={minimum size=0.9cm, draw,circle},
  blank/.style={draw=none},
  edge from parent/.style=
  {draw,edge from parent path={(\tikzparentnode) -- (\tikzchildnode)}},
  level distance = 1.2cm, 
  sibling distance = 0.5cm
} \usetikzlibrary{arrows.meta} \usetikzlibrary{calc} \usetikzlibrary{intersections}

\usepackage{subcaption} \graphicspath{{./}{../figures/}} \usepackage{lipsum}
\usepackage{color}

\begin{document}
\begin{frontmatter}

\title{Generating General Preferential Attachment Networks with \textsf{R} Package \pkg{wdnet}}

\author[1]{
  \inits{Y.}
  \fnms{Yelie}
  \snm{Yuan}  \thanksref{1}  \ead{yelie.yuan@uconn.edu}}
\author[2]{
  \inits{T.}
  \fnms{Tiandong}
  \snm{Wang}}
\author[1]{
  \inits{J.}
  \fnms{Jun}
  \snm{Yan}}
\author[3]{
  \inits{P.}
  \fnms{Panpan}
  \snm{Zhang}}

\thankstext[type=corresp,id=1]{Corresponding author}
\address[1]{Department of Statistics, 
  \institution{University of Connecticut}, \cny{Storrs, Connecticut, USA}}
\address[2]{Shanghai Center for Mathematical Sciences, 
  \institution{Fudan University}, \cny{Shanghai, China}}
\address[3]{Department of Biostatistics, 
  \institution{Vanderbilt University Medical Center}, \cny{Nashville, Tennessee, USA}}

\begin{abstract}
Preferential attachment (PA) network models have a wide range of applications in various scientific disciplines. Efficient generation of large-scale PA networks helps uncover their structural properties and facilitate the development of associated analytical methodologies. Existing software packages only provide limited functions for this purpose with restricted configurations and efficiency. We present a generic, user-friendly implementation of weighted, directed PA network generation with \proglang{R} package \pkg{wdnet}. The core algorithm is based on an efficient binary tree approach. The package further allows adding multiple edges at a time, heterogeneous reciprocal edges, and user-specified preference functions. The engine under the hood is implemented in \proglang{C++}. Usages of the package are illustrated with detailed explanation. A benchmark study shows that \pkg{wdnet} is efficient for generating general PA networks not available in other packages. In restricted settings that can be handled by existing packages, \pkg{wdnet} provides comparable efficiency.
\end{abstract}


\end{frontmatter}

\section{Introduction}
\label{sec:intro}

Preferential attachment (PA) networks are important network models
in scientific research. The
standard PA model~\citep{Barabasi1999emergence} evolves under the
mechanism that a new node is attached to an existing node with probability
proportional to its degree. With the increasing needs of
accommodating the
heterogeneity and complexity of modern networks, a
variety of extended PA network models have been proposed. Examples are
directed PA models~\citep{Bollobas2003directed}, generalized directed
PA models \citep{Britton2020directed},
weighted PA models~\citep{Barrat2004weighted},
and PA models with reciprocal
edges~\citep{Britton2020directed, Wang2022asymptotic, Wang2022random}.
In a general setting, the probability that a node
gets a new edge is proportional to a preference function of some (node-specific)
characteristics (e.g., node degree or strength). Due to their
versatility, PA
models have found a wide range of applications such as friendship
networks~\citep{Momeni2015measuring},
scientific collaboration networks~\citep{Abbasi2012betweenness},
Wikipedia networks~\citep{Capocci2006preferential},
and the World Wide Web~\citep{Kong2008experience}, among others.
Many of these networks are massive in scale.

Efficient generation of large-scale PA networks is critical to the
investigations of their complex local and asymptotic properties. When the
preference function is linear in node degree, \citet{Wan2017fitting} developed a
structured algorithm with complexity \(O(n)\) for generating directed PA networks,
where \(n\) is the number of generation steps. When the preference function is
nonlinear in node degree, however, a naive extension of this algorithm requires
visiting existing nodes one after
another at each sampling step, leading to an increase in complexity to
\(O(n^2)\). Other node-degree-based techniques like stratified
sampling or grouping~\citep{Hadian2016roll} cannot handle continuous edge
weights. An algorithm based on a binary tree~\citep{Atwood2015efficient}
has complexity \(O(n \log n)\) at the cost of
additional storage of subtree information for each node. This algorithm is
promising in handling weighted, directed PA networks with
general preference functions. No user-friendly software
package, however, has been available beyond the \proglang{C}
implementation of \citet{Atwood2015efficient}.

Existing software packages only provide limited functions for PA
network generation. \proglang{Python} package
\pkg{NetworkX}~\citep{Hagberg2009exploring} has a
utility function for generating unweighted, undirected, linear PA
networks. \proglang{R} packages
\pkg{igraph}~\citep{Csardi2006igraph}, \pkg{PAFit}~\citep{Pham2020pafit}, and
\pkg{fastnet}~\citep{Dong2020fastnet} contain functions for generating directed
and/or undirected PA networks, but none of them allows edge weights. Both
\pkg{igraph} and \pkg{PAFit} provide functions for preference functions that are
not linear in node degrees, but they only cover a small class of power and
logarithm functions. Further, no existing package implements the recently
proposed PA models with reciprocity~\citep{Britton2020directed,
  Wang2022asymptotic, Wang2022random}.
See Table~\ref{table:packages} for a brief summary of the functions for
generating PA networks in these packages.

\begin{table}
  \caption{Summary of packages generating PA networks.}
  \label{table:packages}
  \begin{tabularx}{\columnwidth}{@{}lccccX@{}}
    \toprule
    Package & \rotatebox{90}{Undirected} & \rotatebox{90}{Directed} & 
      \rotatebox{90}{Weighted} & \rotatebox{90}{Reciprocal} & 
      Preference function \\
    \midrule
    \pkg{fastnet} & $\checkmark$ & $\checkmark$ & & & 
      Node degree \\
    \pkg{igraph}  & $\checkmark$ & $\checkmark$ & & & 
      Power of node degree plus a positive constant \\
    \pkg{NetworkX}& $\checkmark$ & & & & 
      Node degree \\
    \pkg{PAFit}   & & $\checkmark$ & & & 
      Power or logarithm of node in-degree \\
    \pkg{wdnet}   & $\checkmark$ & $\checkmark$ & $\checkmark$ & $\checkmark$ & 
      General (user-specified) function of node degree/strength \\
    \bottomrule
  \end{tabularx}
\end{table}

We introduce an \proglang{R} package \pkg{wdnet}~\citep{Rpkg:wdnet}
for efficient generations of a general class of PA networks. The core
algorithm is a generalization of the binary tree
approach~\citep{Atwood2015efficient}. Our package contains substantial improvements
in the flexibility for the generation of PA networks: It not only allows
directed edges and edge weights, but also has additional features such as
multiple edge additions, user-defined preference functions, and
heterogeneous reciprocal edges, among others. See Table~\ref{table:packages}
for a summary of the features in comparison with existing packages. The
engine under the hood is implemented in \proglang{C++} for fast speed
and then interfaced to \proglang{R} as facilitated by package
\pkg{Rcpp}~\citep{Eddelbuettel2011rcpp}.

The rest of the paper is organized as follows. In
Section~\ref{sec:algorithm}, we introduce the preliminaries of weighted,
directed PA networks and present the core binary tree algorithm for
generating PA networks with basic configurations.
In Section~\ref{sec:interface}, we illustrate the usage of the main
generation
function and how to control the PA network configurations for advanced
features like adding multiple edges and reciprocal edges, and
defining user-specified preference functions. Performance
comparisons are conducted in
Section~\ref{sec:benchmarks}. Section~\ref{sec:discussion} concludes with a
summary of the paper and a brief introduction of other functions beyond PA network
generation in package \pkg{wdnet}.

\section{Generating weighted, directed PA networks}
\label{sec:algorithm}

We begin with an introduction to weighted,
directed
PA networks and a generic PA network generation framework in
Section~\ref{sec:pa}. The core of an efficient PA network
generation algorithm in package \pkg{wdnet}
is specified in Section~\ref{sec:tree}.

\subsection{Preliminaries}
\label{sec:pa}

For discrete time \(t = 0, 1, 2, \ldots\), let \(G(t) := (V(t), E(t))\)
be a weighted, directed network with
node set \(V(t)\) and edge set \(E(t)\). For any \(v_j, v_k \in V(t)\), let
\((v_j, v_k, w_{jk}) \in E(t)\) denote a directed edge from \(v_j\) to \(v_k\),
where \(w_{jk} > 0\) represents its weight. There can be more than one
edges
from \(v_j\) to \(v_k\). For the special case of
\(j = k\), \((v_j, v_k, w_{jk}) \in E(t)\) is a self-loop. By
convention, an initial (or seed) network \(G(0)\) has at least one
node and one edge.

We consider weighted, directed PA networks that allow adding
multiple edges at each epoch. For illustration, we begin with a
standard directed PA network that adds one edge
at a time for now. There are three edge creation scenarios, respectively
associated with probabilities \(\alpha, \beta, \gamma \geq 0\), subject to
\(\alpha + \beta + \gamma = 1\).
Note that we do not allow \(\beta = 1\) to avoid
degenerative situations. At each step \(t \ge 1\), we flip a three-sided coin
whose outcomes correspond to the three edge creation scenarios as follows:

\begin{enumerate}
  \item[(1)] With probability $\alpha$, add
  a new edge from a new node to an existing one from $G(t - 1)$;
  \item[(2)] With probability $\beta$, add a
  new edge between two existing nodes from $G(t - 1)$ (self-loops 
  are allowed);
  \item[(3)] With probability $\gamma$, add
  a new edge from an existing node from $G(t - 1)$ to a new one.
\end{enumerate}

\begin{figure}[tbp]
  \begin{center}
        \begin{tikzpicture}
            \draw[fill = gray!20] (-5, 0) ellipse (1.6 and 1) ;
            \draw (-4.2, 0) node[draw = black, circle, minimum
            size =
            0.8cm, fill = blue!20] (j1) {$v_j$} ;
            \draw (-3.5, -1.5) node[draw = black, circle,
            minimum
            size = 0.8cm] (i1)  {$v_i$} ;
            \draw[-latex, thick] (i1) -- (j1) ;
            \draw[fill = gray!20] (0, 0) ellipse (1.6 and 1) ;
            \draw (-0.8, 0) node[draw = black, circle, minimum
            size =
            0.8cm, fill = blue!20] (i2) {$v_i$} ;
            \draw (0.8, 0) node[draw = black, circle, minimum
            size = 0.8cm, fill = blue!20] (j2)  {$v_j$} ;
            \draw[-latex, thick] (i2) -- (j2) ;
            \draw[fill = gray!20] (5, 0) ellipse (1.6 and 1) ;
            \draw (4.2, 0) node[draw = black, circle, minimum
            size =
            0.8cm, fill = blue!20] (i3) {$v_i$} ;
            \draw (3.5, -1.5) node[draw = black, circle, minimum
            size = 0.8cm] (j3)  {$v_j$} ;
            \draw[-latex, thick] (i3) -- (j3) ;
        \end{tikzpicture}
        \caption{Three edge creation scenarios corresponding to
            $\alpha$, $\beta$ and $\gamma$ schemes (from left to right),
            respectively.}
        \label{fig:PAscenario}
    \end{center}
\end{figure}
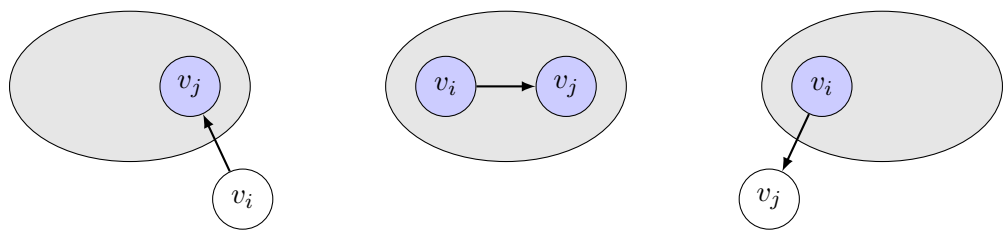

For convenience, we call these three scenarios \(\alpha\), \(\beta\),
and \(\gamma\) schemes, respectively; see Figure~\ref{fig:PAscenario} for a
graphical illustration.

Once an edge creation scenario is decided, we need to
determine the corresponding source and/or target nodes. The
probability
of each candidate node from the current network being selected is
proportional to its preference score, which is given by a function
(called preference function) of node-specific characteristics. For
unweighted, directed PA networks, the most commonly used
characteristics are
out- and in-degrees, whereas for weighted, directed PA networks,
out- and
in-strengths are usually adopted. Let
\[
  \text{O}(v_j, t) := \sum_{k:(v_j, v_k, w_{jk}) \in E(t)} w_{jk} 
  \qquad \mbox{and} 
  \qquad \text{I}(v_j, t) := \sum_{k:(v_k, v_j, w_{kj}) \in E(t)} w_{kj}
\]
represent the out- and in-strength of \(v_j \in V(t)\), respectively.
Let \(\theta_1(v_j, t) := f_1(\text{O}(v_j, t), \text{I}(v_j,t))\)
be the preference score for sampling \(v_j\) as a
source node for a newly added edge at step \(t + 1\), with a non-negative
function \(f_{1}(\cdot)\) called source preference function. Then the
probability
of node \(v_j \in V(t)\) being selected as a source node at time \(t + 1\)
is given by
\begin{equation*}
  \frac{\theta_1(v_j, t)}{\sum_{v_k \in V(t)} \theta_1(v_k, t)}.
\end{equation*}
Similarly, with a non-negative target preference function \(f_{2}(\cdot)\), one
can define the preference score for sampling \(v_j \in V(t)\) as a
target node at time \(t + 1\) as well as the associated sampling
probability. The default option for both \(f_1\) and \(f_2\) in the package is
a power function:
\begin{equation}
  \label{eq:fxy}
  f(x, y) := a_1 x^{a_2} + a_3 y^{a_4} + a_5,
\end{equation}
where the parameters, \(a_i\), \(i = 1, \ldots, 5\), are specified by the
users and can be different for \(f_1\) and \(f_2\).
User-defined preference functions are also allowed;
see Section~\ref{sec:interface_features} for details.

Once the source and target nodes of a new edge are selected,
its weight is drawn independently from a
distribution with probability density or mass
function \(h\) on a positive support. The
in- and out-strengths of the corresponding nodes are also
updated, as well as their source and target preference scores. Then the
algorithm proceeds to the next step.

\begin{algorithm}[tbp]
    \caption{Generating a weighted, directed PA network.}
    \label{alg:PA}
  \SetNoFillComment
  \KwIn{Number of steps $n$; \newline
    initial network $G(0) = (V(0), E(0))$; \newline
    probabilities for three edge creation scenarios $\alpha, \beta$, 
    $\gamma$; \newline
    probability density (or mass) function $h$ for drawing edge weights 
    from; \newline
    preference functions for source node $f_{1}$ and for target node $f_{2}$.}
    \KwOut{$G(n) = (V(n), E(n))$.}
    \SetKwProg{Fn}{Function}{:}{}
    \SetKwFunction{SampleNode}{Sample\_Node}
    \SetKwProg{Algr}{Algorithm}{:}{}
    \Algr{}{
    Initialize $(n + |V(0)|)$-dimensional zero vectors of out- and in-strengths, 
    $\bm{\text{O}}$ and $\bm{\text{I}}$\;
    Initialize $(n + |V(0)|)$-dimensional zero vectors of source and target preference 
    scores $\bm{\theta}_1$ and $\bm{\theta}_2$\;
    Update $\bm{\text{O}}$, $\bm{\text{I}}$, $\bm{\theta}_1$ and $\bm{\theta}_2$ 
      with initial network $G(0)$\;
    $t \leftarrow 1$\;
    \While{$t \leq n$}{
      $N \leftarrow |V(t - 1)|$
      \tcc*{Number of nodes in $G(t - 1)$}
      Draw $\psi \sim \mathrm{Unif}(0, 1)$\;
      \uIf(\tcc*[f]{$\alpha$ scheme}){$\psi \leq \alpha$}{
        $j \leftarrow N + 1$ 
        \tcc*{Source node index}
        $k \leftarrow$ \SampleNode{$V(t - 1), 2$} 
        \tcc*{Target node index; 
        the "2" as the second argument indicates target node sampling.}
        $V(t) \leftarrow V(t - 1) \cup \{v_j$\}\;
      }
      \uElseIf(\tcc*[f]{$\beta$ scheme}){$\alpha < \psi \leq \alpha + \beta$}{
        $j \leftarrow$ \SampleNode{$V(t - 1), 1$}
        \tcc*{The "1" as the second argument indicates source node sampling}
        $k \leftarrow$ \SampleNode{$V(t - 1), 2$}\;
      }
      \uElseIf(\tcc*[f]{$\gamma$ scheme}){$\psi > \alpha + \beta$}{
        $j \leftarrow$ \SampleNode{$V(t - 1), 1$}\;
        $k \leftarrow N + 1$\;
        $V(t) \leftarrow V(t - 1) \cup \{v_k\}$\;
      }
      Draw $w$ from weight distribution $h$
      \tcc*{Sample edge weight}
      $E(t) \leftarrow E(t - 1) \cup \left\{(v_j, v_k, w)\right\}$
      \tcc*{Add the new edge to $G(t)$}
      $\bm{\text{O}}[j] \leftarrow \bm{\text{O}}[j] + w$
      \tcc*{Update preference function inputs}
      $\bm{\text{I}}[k] \leftarrow \bm{\text{I}}[k] + w$\;
      $\bm{\theta}_1[j] \leftarrow f_{1}(\bm{\text{O}}[j], \bm{\text{I}}[j])$
      \tcc*{Update preference scores}
      $\bm{\theta}_2[j] \leftarrow f_{2}(\bm{\text{O}}[j], \bm{\text{I}}[j])$\;
      $\bm{\theta}_1[k] \leftarrow f_{1}(\bm{\text{O}}[k], \bm{\text{I}}[k])$\;
      $\bm{\theta}_2[k] \leftarrow f_{2}(\bm{\text{O}}[k], \bm{\text{I}}[k])$\;
      $t \leftarrow t + 1$\;
    }
    \KwRet{$G(n) = (V(n), E(n))$}\;
  }
\end{algorithm}

Algorithm~\ref{alg:PA} summarizes the core structure of generating a
weighted, directed PA network. The bottleneck of the
algorithm is how to efficiently sample source or target nodes, i.e.,
the \code{Sample\_Node()} function in Algorithm~\ref{alg:PA}. We use
the sampling procedure for source nodes as an illustration.
At time \(t + 1\), the sampling step takes an updated
vector of preference scores \(\{\theta_1(v_j, t): v_j \in V(t)\}\) as input.
In fact, the task is straightforward. Given the
grid of increasing breakpoints formed by cumulative sums of the current
preference scores,
find an appropriate interval that contains a uniform random variable
\(U\) drawn from
{\rm Unif}\((0, \sum_{v_j \in V(t)} \theta_1(v_j, t))\).
This can be done by sequentially subtracting node preference
scores from \(U\) until we find the node such that removing its
preference score would cause \(U \leq 0\). The above sampling method
is a fundamental linear search, as it has to visit each of the
existing nodes (one after another), and keeps updating their
preference scores.
This sampling approach is the \code{linear} method in the package, and the
complexity of network generation by using this method is \(O(n^2)\).

Fast sampling is possible for some special cases like when source
and target preference functions are linear in node out- and
in-degrees, respectively. Without loss of generality, consider a
source preference function in the form of
\(f_{1}(x, y) = x + a_5\). When the edges are unweighted, the
interval (containing \(U\)) can be determined by one
uniform draw~\citep{Wan2017fitting}. This algorithm acts like by
putting the node labels into a bag the same number of times as their
out-degrees and then drawing a label from the bag, which is
analogous to the P\'{o}lya urn theory~\citep{Mahmoud2008polya}.
This technique
is called \code{bag} in package \pkg{igraph}, and the same name
is adopted in package \pkg{wdnet}. When the edges are weighted, by
a clever maneuver, the sampling step for the whole generation
process can be done in one batch with a pre-set cumulative sum
vector of the edge weights by using the base \proglang{R} function
\code{findInterval()}.
This is an extension of the \code{bag} algorithm, so we named it
\code{bagx}. See Appendix~\ref{appendix:alternative} for more
details about \code{bag} and \code{bagx}.

\subsection{Node sampling based on a binary tree}
\label{sec:tree}

Our recommended approach for \code{Sample\_Node()}
is a binary tree approach that extends the
algorithm in \citet{Atwood2015efficient} to weighted, directed PA
networks with general preference functions. In a binary tree
structure, each node has no more than two
children nodes. The two children nodes of a parent node are
distinguished by their positions, i.e., the left and the right
child. Except for the root node, each node has only one parent
node. A complete binary tree refers to a binary tree with all levels
fully filled except for the last level. The last level is not
necessarily completely filled, but has to be filled from left to
right. A hypothetical example of complete binary tree is given in
Figure~\ref{fig:tree_only}.
An important application of binary trees is searching. The complexity
of searching a specific node in a binary tree with \(n\) nodes is of
order \(O(\log n)\)~\citep{Mahmoud1992evolution}, which is more
efficient than the linear search of complexity \(O(n)\).

\begin{figure}[tbp]
  \begin{subfigure}[t]{0.43\linewidth}
    \centering
    \begin{tikzpicture}
      \draw (-0.9, 0) node[draw = blue, circle, minimum size = 
      0.9cm, fill = blue, opacity = 0.2] {};
      \draw (-0.9, 0) node[draw = blue, circle, minimum size = 
      0.9cm]
      (v1) {$v_1$};
      \draw (0.9, 0) node[draw = blue, circle, minimum size = 
      0.9cm, fill = blue, opacity = 0.2] {};
      \draw (0.9, 0) node[draw = blue, circle, minimum size = 0.9cm]
      (v2) {$v_2$};
      \draw (2.7, 0) node[draw = black, circle, minimum size = 0.9cm]
      (v3) {$v_3$};
      \draw (-0.9, 1.8) node[draw = black, circle, minimum size = 0.9cm]
      (v4) {$v_4$};
      \draw (-2.7, 0) node[draw = black, circle, minimum size = 0.9cm]
      (v5) {$v_5$};
      \draw (-2.7, 1.8) node[draw = black, circle, minimum size = 0.9cm]
      (v6) {$v_6$};
      \draw (-0.9, -1.8) node[draw = black, circle, minimum size = 0.9cm]
      (v7) {$v_7$};
      \draw (2.7, 1.8) node[draw = black, circle, minimum size = 0.9cm]
      (v8) {$v_8$};
      \draw (0.9, -1.8) node[draw = black, circle, minimum size = 0.9cm]
      (v9) {$v_9$};
      \draw (0.9, 1.8) node[draw = black, circle, minimum size = 0.9cm]
      (v10) {$v_{10}$};
      \draw[-latex, thick, blue] (v1) -- node [midway, above] {1} (v2);
      \draw[-latex, thick] (v3) -- node [midway, above] {2} (v2);
      \draw[-latex, thick] (v1) -- node [midway, right] {1} (v4);
      \draw[-latex, thick] (v5) -- node [midway, above] {2} (v1);
      \draw[-latex, thick] (v6) -- node [midway, above] {3} (v4);
      \draw[-latex, thick] (v7) -- node [midway, right] {2} (v1);
      \draw[-latex, thick] (v8) -- node [midway, right] {1} (v3);
      \draw[-latex, thick] (v9) -- node [midway, right] {2} (v2);
      \draw[-latex, thick] (v10) -- node [midway, above] {1} (v4);
    \end{tikzpicture}
    \caption{A simple network. Edge weights are marked next to the edges.}
    \label{fig:network_only}
  \end{subfigure}
  \begin{subfigure}[t]{0.55\linewidth}
    \centering
    \begin{tikzpicture}
      \Tree
      [.\node (v1) {$v_1$};
        [.\node (v2) {$v_2$};
          [.\node (v4) {$v_4$};
            \node (v8) {$v_8$};
            \node (v9) {$v_9$};
          ]
          [.\node (v5) {$v_5$};
            \node (v10) {$v_{10}$};
            \edge[blank]; \node[blank]{};
          ]
        ]
        [.\node (v3) {$v_3$};
          \node (v6) {$v_6$};
          \node (v7) {$v_7$};
        ]
      ]
    \end{tikzpicture}
    \caption{The binary tree structure.}
    \label{fig:tree_only}
  \end{subfigure}
  \\[0.2in]
  \begin{subfigure}{\linewidth}
    \centering
    \begin{tabular}{@{}*{10}{c}@{}}
      \toprule
      Node & $\kappa(v_j)$ & $l(v_j)$ & $r(v_j)$
      & $\text{O}(v_j)$ & $\text{I}(v_j)$ & $\theta_1(v_j)$ & $\theta_2(v_j)$
      & $\eta_1(v_j)$ & $\eta_2(v_j)$ \\
      \midrule
      $v_1$  &   /   & $v_2$ & $v_3$ & 2 & 4 & 3 & 5 & 25& 25\\
      $v_2$  & $v_1$ & $v_4$ & $v_5$ & 0 & 5 & 1 & 6 & 12& 16\\
      $v_3$  & $v_1$ & $v_6$ & $v_7$ & 2 & 1 & 3 & 2 & 10 & 4 \\
      $v_4$  & $v_2$ & $v_8$ & $v_9$ & 0 & 5 & 1 & 6 & 6 & 8 \\
      $v_5$  & $v_2$ & $v_{10}$ & /  & 2 & 0 & 3 & 1 & 5 & 2 \\
      $v_6$  & $v_3$ &   /   &   /   & 3 & 0 & 4 & 1 & 4 & 1 \\
      $v_7$  & $v_3$ &   /   &   /   & 2 & 0 & 3 & 1 & 3 & 1 \\
      $v_8$  & $v_4$ &   /   &   /   & 1 & 0 & 2 & 1 & 2 & 1 \\
      $v_9$  & $v_4$ &   /   &   /   & 2 & 0 & 3 & 1 & 3 & 1 \\
      $v_{10}$&$v_5$ &   /   &   /   & 1 & 0 & 2 & 1 & 2 & 1 \\
      \bottomrule
    \end{tabular}
    \caption{Node attributes.}
    \label{fig:node_attaributes}
  \end{subfigure}
  \caption{A generated network with initial graph colored with 
  blue (panel a), its corresponding complete binary tree structure 
  (panel b) and a summary of node attributes (panel c); the source and
  target preference functions are respectively given by
  $f_{1}(x, y) = x + 1$ and $f_{2}(x, y) = y + 1$.}
  \label{fig:binary_tree}
\end{figure}

We translate a PA network to a binary tree as follows. Each
node in a PA network corresponds to a node in the associated
complete binary tree based on the time of its creation. Suppose that
\(G(0)\) contains only one node
\(v_1\) with a self-loop, then \(v_1\) becomes the root of the complete
binary tree. Then node \(v_2\) that joins the PA network at \(t = 1\)
is the left child of \(v_1\), and the next new node \(v_3\), which joins
the PA network at \(t = 2\), is the right child of \(v_1\). For the next
newcomer \(v_4\) (at time \(t = 3\)), it is attached to \(v_2\) as a left
child since the first level (consisting of \(v_2\) and \(v_3\)) is fully
filled. The transition continues in this fashion until all nodes in
the PA network are added to the complete binary tree. For an initial
graph \(G(0)\) containing more than one nodes, a node enumeration
\(\{v_1, v_2, \ldots, v_{|V(0)|}\}\) is required before constructing the
binary tree; see Figures~\ref{fig:network_only}
and~\ref{fig:tree_only} for an example with an initial graph consisting of
two nodes (connected by one edge which is colored with blue).
Different enumeration orders of the initial network result in different
binary trees and, consequently, different networks because of the
underlying sampling mechanism.

Having built a complete binary tree, we augment the nodes therein
according to a collection of node attributes. At step \(t\), the
binary tree node \(v_j\) stores the following information:
parent (except for \(v_1\))
\(\kappa(v_j)\), left child \(l(v_j)\), right child
\(r(v_j)\), out-strength \(\text{O}(v_j, t)\), in-strength \(\text{I}(v_j, t)\),
preference score as a source node \(\theta_1(v_j, t)\), preference score as a
target node \(\theta_2(v_j, t)\).
For now we consider \(\theta_1\) and \(\theta_2\) as functions of node
out- and in-strengths, but in general, \(\theta_1\) and \(\theta_2\) can be
functions of any node-level characteristics. Additionally, let
\(\eta_1(v_j, t)\) and \(\eta_2(v_j, t)\) denote the total preference of
source and target nodes of the subtree
(a portion of the binary tree consisting of a node
and all of its descendants) with root \(v_j\),
respectively, giving rise to the following relationship:
\begin{align*}
  \eta_i(v_j, t) &= \eta_i(l(v_j), t) + \eta_i(r(v_j), t) + \theta_i(v_j, t),
  \qquad i \in \{1, 2\}.
\end{align*}
Figure~\ref{fig:node_attaributes} summarizes the node attributes,
including \(\eta_1\) and \(\eta_2\), from the complete binary tree
(in Figure~\ref{fig:tree_only}) that is constructed from the weighted
network in Figure~\ref{fig:network_only}.

\begin{algorithm}[tbp]
  \caption{Node sampling function based on a binary tree storage structure.}
  \label{alg:samplenode}
    \SetKwProg{Fn}{Function}{:}{}
    \SetKwFunction{SampleNode}{Sample\_Node}
    \BlankLine
    \KwIn{Node set $V(t - 1)$; \newline
        $i \in\{1, 2\}$ for sampling a source or a target node, respectively.}
    \KwOut{$j$, the index of the sampled node.}
    \Fn{\SampleNode{$V(t - 1), i$}}{
        $j = 1$
    \tcc*{Start from the root $v_1$}
        Draw $U \sim \mathrm{Unif} (0, \eta_i(v_1, t - 1))$\;
    \While{$j \leq |V(t - 1)|$}{
      $U \leftarrow U - \theta_i(v_j, t - 1)$\;
      ${\tt temp} \leftarrow \eta_i(l(v_j), t - 1)$\;
      \uIf(\tcc*[f]{Search in the subtree with root $l(v_j)$})
        {$0 < U \le {\tt temp}$}{
        $j \leftarrow \mbox{ index of } l(v_j)$\;
      }
      \uElseIf(\tcc*[f]{Search in the subtree with root $r(v_j)$})
        {$U > {\tt temp}$}{
        $U \leftarrow U - {\tt temp}$\;
        $j \leftarrow \mbox{ index of }r(v_j)$\;
      }
      \uElseIf(\tcc*[f]{Return the index of the sampled node $v_j$})
        {$U \le 0$}{
        \KwRet{$j$\;}
      }
    }
  }
\end{algorithm}

Node sampling based on the binary tree structure, available as the
\code{binary} method in \pkg{wdnet}, is summarized in
Algorithm~\ref{alg:samplenode}. Generally, the algorithm searches for the
subtree to which the potentially sampled
node belongs in a recursive manner until the
root of the resulting subtree (or the node itself if it is at the bottom level)
is returned. The subtree-based searching substantially reduces the
complexity of the sampling step from \(O(n)\) (for the linear search algorithm)
to \(O(\log n)\). The network generation algorithm with binary search thus has
complexity \(O(n \log n)\).

Upon the creation of a new edge \((v_j, v_k, w_{jk})\), the following
quantities need to be updated: node strengths \(\text{O}(v_j, t)\) and
\(\text{I}(v_k, t)\); preference
scores \(\theta_1(v_j, t)\), \(\theta_1(v_k, t), \theta_2(v_j, t)\) and
\(\theta_2(v_k, t)\); total preference scores of subtrees \(\eta_1(v_j, t)\),
\(\eta_2(v_j, t)\),
\(\eta_1(\kappa(v_j), t)\), \(\eta_2(\kappa(v_j), t)\), etc.
The update of total preference scores is not shown in the algorithm.
It traces the growth path through subtrees (backwards), and has the
same time complexity \(O(\log n)\) as the sampling method.

\section{Usage}
\label{sec:interface}

We start with the main function to generate PA networks with basic
configurations in Section~\ref{sec:interface_rpanet}, and then
introduce additional features in Section~\ref{sec:interface_features}.

\subsection{Main generation function}
\label{sec:interface_rpanet}

The function \code{rpanet()} is used to generate PA networks.

\begin{Shaded}
\begin{Highlighting}[]
\KeywordTok{library}\NormalTok{(}\StringTok{"wdnet"}\NormalTok{)}
\KeywordTok{args}\NormalTok{(rpanet)}
\end{Highlighting}
\end{Shaded}

\begin{verbatim}
function (nstep, initial.network = list(edgelist = matrix(c(1, 
    2), nrow = 1), edgeweight = 1, directed = TRUE), control, 
    method = c("binary", "linear", "bagx", "bag")) 
NULL
\end{verbatim}

The first three arguments of \code{rpanet()}
are: the number of steps (\code{nstep}),
the initial network (\code{initial.network}), and a list of control
parameters (\code{control}). Specifications of the control parameters
are done through a collection of functions as we proceed.
The \code{method} argument specifies which of the following four
implemented methods is used to generate a PA network:
\code{binary} (default), \code{linear}, \code{bagx}, and \code{bag}.

With respect to the required inputs in Algorithm~\ref{alg:PA}, we
elaborate the usage of \code{rpanet()} and its specifications via
the \code{control} argument as follows.

\paragraph{Initial network}

The \code{initial.network} is specified by a list containing a
matrix of edges (\code{edgelist}) in the order of edge creations,
a vector of edge weights (\code{edgeweight}), and a logical argument
(\code{directed}) indicating whether the initial network
as well as the generated network are directed.
Each row of \code{edgelist} has two elements
specifying the two nodes of an edge. The length of
\code{edgeweight} is equal to the number of rows of \code{edgelist}.
If \code{edgeweight} is not specified, all edges from the initial
network are assumed to have weight 1.
The default initial network has only one edge, \((1, 2, 1.0)\),
corresponding to a network consisting of two nodes with a unit-weight
edge from node~1 to node~2.
The \code{initial.network} can also be specified by a
\code{wdnet} object, which can be constructed via utility functions
\code{edgelist\_to\_wdnet()} or \code{adj\_to\_wdnet()}.
The following example
sets up an initial network with two weighted edges, \((1, 2, 0.5)\)
and \((3, 4, 2.0)\).

\begin{Shaded}
\begin{Highlighting}[]
\NormalTok{netwk0 \textless{}{-}}\StringTok{ }\KeywordTok{list}\NormalTok{(}\DataTypeTok{edgelist =} \KeywordTok{matrix}\NormalTok{(}\KeywordTok{c}\NormalTok{(}\DecValTok{1}\NormalTok{, }\DecValTok{2}\NormalTok{, }\DecValTok{3}\NormalTok{, }\DecValTok{4}\NormalTok{), }\DataTypeTok{nrow =} \DecValTok{2}\NormalTok{, }\DataTypeTok{byrow =} \OtherTok{TRUE}\NormalTok{),}
    \DataTypeTok{edgeweight =} \KeywordTok{c}\NormalTok{(}\FloatTok{0.5}\NormalTok{, }\FloatTok{2.0}\NormalTok{), }\DataTypeTok{directed =} \OtherTok{TRUE}\NormalTok{)}
\end{Highlighting}
\end{Shaded}

\paragraph{Edge scenarios}

The function \code{rpa\_control\_scenario()} is used to specify the
probability of each edge creation scenario.
Based on the real data analysis in \cite{Wan2017fitting},
we also include two additional edge creation scenarios to the
\(\alpha\), \(\beta\) and \(\gamma\) schemes introduced in Section~\ref{sec:pa}:
(1) the \(\xi\) scheme where a new edge is added between two new nodes, and
(2) the \(\rho\) scheme where a new node with a
self-loop is added. Self-loops are allowed in the \(\beta\) scheme by setting
\code{beta.loop} to be \code{TRUE}. When
\code{beta.loop = FALSE},
the order of sampling source and target nodes may affect the
structure of generated PA network as controlled by the logical arguments
\code{source.first}.
The default settings of these arguments are \(\alpha = 1\),
\(\beta = \gamma = \xi = \rho = 0\),
\code{beta.loop = source.first = TRUE}.
The following example sets up a configuration that excludes self-loops
under the \(\beta\) scheme and samples target nodes before source
nodes.

\begin{Shaded}
\begin{Highlighting}[]
\NormalTok{ctr1 \textless{}{-}}\StringTok{ }\KeywordTok{rpa\_control\_scenario}\NormalTok{(}\DataTypeTok{alpha =} \FloatTok{0.2}\NormalTok{, }\DataTypeTok{beta =} \FloatTok{0.6}\NormalTok{, }\DataTypeTok{gamma =} \FloatTok{0.2}\NormalTok{,}
    \DataTypeTok{beta.loop =} \OtherTok{FALSE}\NormalTok{, }\DataTypeTok{source.first =} \OtherTok{FALSE}\NormalTok{)}
\end{Highlighting}
\end{Shaded}

\paragraph{Edge weights}

Edge weights are controlled by \code{rpa\_control\_edgeweight()} through its
\code{sampler} argument. This argument accepts a function that takes a single
parameter, representing the number of sampled values, and returns a vector of
sampled edge weights. Note that the sampled values must be positive real
numbers. The default setting is \code{sampler = NULL}, referring to the case
where all new edges take unit
weight. As shown in the following example, edge weights are sampled
from a gamma distribution with shape~5 and scale~0.2; the ``\code{+}'\,' operator
has been overloaded to concatenate multiple control lists.

\begin{Shaded}
\begin{Highlighting}[]
\NormalTok{my\_rgamma \textless{}{-}}\StringTok{ }\ControlFlowTok{function}\NormalTok{(n) }\KeywordTok{rgamma}\NormalTok{(n, }\DataTypeTok{shape =} \DecValTok{5}\NormalTok{, }\DataTypeTok{scale =} \FloatTok{0.2}\NormalTok{)}
\NormalTok{ctr2 \textless{}{-}}\StringTok{ }\NormalTok{ctr1 }\OperatorTok{+}\StringTok{ }\KeywordTok{rpa\_control\_edgeweight}\NormalTok{(}\DataTypeTok{sampler =}\NormalTok{ my\_rgamma)}
\end{Highlighting}
\end{Shaded}

\paragraph{Preference functions}

The default preference function is in the form of \(f(x, y)\) given in
Equation~\eqref{eq:fxy}, which covers a wide range of sub-linear, linear and
super-linear functions. The function \code{rpa\_control\_preference()}
controls the
configuration of this \(f(x, y)\) with \code{ftype = "default"} along with two
arguments \code{sparams} and \code{tparams} which specify the parameters of the
source and target preference functions of Equation~\eqref{eq:fxy}, respectively.
For directed PA networks, the default source and target preference
functions are,
respectively, \(f_{1}(x, y) = x + 1\) and \(f_{2}(x, y) = y + 1\).
This is controlled by default value of \code{sparams = c(1, 1, 0, 0, 1)} and
\code{tparams = c(0, 0, 1, 1, 1)}.
The following example sets the source preference function to
\(f_{1}(x, y) = x^2 + 1\) and the target preference function to
\(f_{2}(x, y) = y^2 + 1\):

\begin{Shaded}
\begin{Highlighting}[]
\NormalTok{ctr3 \textless{}{-}}\StringTok{ }\NormalTok{ctr2 }\OperatorTok{+}\StringTok{ }\KeywordTok{rpa\_control\_preference}\NormalTok{(}\DataTypeTok{ftype =} \StringTok{"default"}\NormalTok{, }
  \DataTypeTok{sparams =} \KeywordTok{c}\NormalTok{(}\DecValTok{1}\NormalTok{, }\DecValTok{2}\NormalTok{, }\DecValTok{0}\NormalTok{, }\DecValTok{0}\NormalTok{, }\DecValTok{1}\NormalTok{), }\DataTypeTok{tparams =} \KeywordTok{c}\NormalTok{(}\DecValTok{0}\NormalTok{, }\DecValTok{0}\NormalTok{, }\DecValTok{1}\NormalTok{, }\DecValTok{2}\NormalTok{, }\DecValTok{1}\NormalTok{))}
\end{Highlighting}
\end{Shaded}

For undirected networks, the default preference function has the form
\(g(x) = x^{b_1} + b_2\), and argument \code{params} specifies the
preference parameters \(b_1\) and \(b_2\), with default values given by
\(b_1 = b_2 = 1\).

We further allow users to specify their own preference
functions; see Section~\ref{sec:interface_features} for details.

\paragraph{Returned value}

Function \code{rpanet()} returns a list of class \code{wdnet} containing the
following components: \code{newedge} is a vector summarizing the number of new
edges added at each step; \code{edge.attr} is a data frame containing edge
weights and edge creation scenarios,
where edges from schemes \(\alpha\), \(\beta\), \(\gamma\), \(\xi\), \(\rho\)
are respectively denoted as scenarios 1, 2, 3, 4, 5,
and edges from the initial network are denoted as scenario 0;
\code{node.attr} is a data frame containing
node out- and in-strengths as well as source and target preference scores. Other
items are self-explanatory.

\begin{Shaded}
\begin{Highlighting}[]
\KeywordTok{set.seed}\NormalTok{(}\DecValTok{12}\NormalTok{)}
\NormalTok{netwk3 \textless{}{-}}\StringTok{ }\KeywordTok{rpanet}\NormalTok{(}\DataTypeTok{nstep =} \FloatTok{1e3}\NormalTok{, }\DataTypeTok{initial.network =}\NormalTok{ netwk0, }\DataTypeTok{control =}\NormalTok{ ctr3)}
\KeywordTok{names}\NormalTok{(netwk3)}
\end{Highlighting}
\end{Shaded}

\begin{verbatim}
[1] "edgelist"  "newedge"   "control"   "directed"  "edge.attr" "weighted" 
[7] "node.attr"
\end{verbatim}

\begin{Shaded}
\begin{Highlighting}[]
\KeywordTok{print}\NormalTok{(netwk3)}
\end{Highlighting}
\end{Shaded}

\begin{verbatim}
Weighted: TRUE
Directed: TRUE
Number of edges: 1002
Number of nodes: 402

Edges:
  source target    weight scenario
1      1      2 0.5000000        0
2      3      4 2.0000000        0
3      5      2 0.4591485        1
4      5      6 0.4347588        3
5      5      7 0.6565225        3
...omitted remaining edges

Node attributes:
      outs        ins     spref       tpref
1 4.523894  0.0000000 21.465613    1.000000
2 0.000000  1.7270612  1.000000    3.982741
3 2.000000  0.8624366  5.000000    1.743797
4 2.823930  2.0000000  8.974579    5.000000
5 2.386927 43.7566871  6.697419 1915.647667
...omitted remaining nodes
\end{verbatim}

\subsection{Additional features}
\label{sec:interface_features}

The package \pkg{wdnet} provides a few additional distinctive
features in the PA network generation process that are not available in other
software packages. These features are obtained by adapting the
Algorithm~\ref{alg:PA}.

\paragraph{Multiple edge addition}

The creation of multiple edges at one step is controlled by function
\code{rpa\_control\_newedge()}. The first argument of this function,
\code{sampler}, determines the distribution of the number of new edges to be
added in the same step. This argument accepts a function that takes a single
parameter, representing the number of values to be sampled, and returns a vector
of sampled number of new edges. Note that the sampled values must be positive
integers. By default, \code{sampler} is set to \code{NULL}, representing the
addition of only one edge at each step.

When more than one edges are added at one step, we keep the node strengths and
their preference scores unchanged until all edges at this step have been added.
Users need to specify whether to sample the candidate nodes with
replacements or not. For directed networks,
the logical arguments \code{snode.replace} and \code{tnode.replace}
determine whether the source and target nodes are sampled
with replacement, respectively. For undirected networks,
only one logical argument \code{node.replace} needs to be specified.

The code below updates the setting from \code{ctr3} by letting
the number of new edges follows a unit-shifted Poisson
distribution~\citep{Wang2023poisson} with probability mass function
\[
  \Pr(X = k) = e^{-2} \frac{2^{k - 1}}{(k - 1)!}, \qquad
  k \geq 1.
\]
Both source and target nodes are sampled \emph{without} replacement.

\begin{Shaded}
\begin{Highlighting}[]
\NormalTok{ctr4 \textless{}{-}}\StringTok{ }\NormalTok{ctr3 }\OperatorTok{+}\StringTok{ }\KeywordTok{rpa\_control\_newedge}\NormalTok{(}\DataTypeTok{sampler =} \ControlFlowTok{function}\NormalTok{(n) }\KeywordTok{rpois}\NormalTok{(n, }\DecValTok{2}\NormalTok{) }\OperatorTok{+}\StringTok{ }\DecValTok{1}\NormalTok{,}
    \DataTypeTok{snode.replace =} \OtherTok{FALSE}\NormalTok{, }\DataTypeTok{tnode.replace =} \OtherTok{FALSE}\NormalTok{)}
\end{Highlighting}
\end{Shaded}

\paragraph{Reciprocal edges}

Reciprocal edges are mutual links between two nodes.
We allow reciprocal edges under a heterogeneous
setting~\citep{Wang2022random} where each node belongs to one of the \(K\ge 1\)
groups. With the emergence of each new node, its
group label is given according to a user-specified probability vector
\(\bm{\pi} := (\pi_1, \pi_2, \ldots, \pi_K)\), where \(0 \leq \pi_{k} \leq 1\)
represents the probability that the node belongs to group
\(k \in \{1, 2, \ldots, K\}\).
Similar to stochastic block models, there is a probability block
matrix
\(\bm{q} := (q_{k\ell})_{K \times K}\) (not necessarily symmetric),
which is also specified by the users, to determine the probability of adding a
reciprocal edge for each new edge joining the network. For example, consider a new edge
\((v_i, v_j, w_{ij})\) where \(v_i\) and \(v_j\) are
respectively
labeled with \(k = 2\) and \(\ell = 3\), then its reciprocal correspondence
\((v_j, v_i, w_{ji})\) is added to the network instantaneously with probability
\(q_{32}\).
The weight of the reciprocal edge (if added), \(w_{ji}\), is independently sampled
with configurations specified in \code{rpa\_control\_edgeweight()}.
When more than one new edges are added at a step,
the reciprocal edge for each of them is added independently, one after
another.

The function \code{rpa\_control\_reciprocal()} gives the
configurations of reciprocal
edges. The arguments \code{group.prob} and \code{recip.prob} specify
the probability vector \(\bm{\pi}\) and the block probability matrix
\(\bm{q}\), respectively. In
addition, the logical argument \code{selfloop.recip} determines
whether reciprocal edges for self-loops are allowed. Their
default settings are
\code{group.prob = NULL}, \code{recip.prob = NULL} and
\code{selfloop.recip = FALSE}, respectively, referring to the case of no
immediate reciprocal edges. The following example creates a configuration with
\(\bm{\pi} = (0.4, 0.6)\) and
\begin{align*}
  \bm{q} = 
  \begin{pmatrix}
    0.4 & 0.1 \\
    0.2 & 0.5
  \end{pmatrix}.
\end{align*}

\begin{Shaded}
\begin{Highlighting}[]
\NormalTok{ctr5 \textless{}{-}}\StringTok{ }\NormalTok{ctr4 }\OperatorTok{+}\StringTok{ }\KeywordTok{rpa\_control\_reciprocal}\NormalTok{(}\DataTypeTok{group.prob =} \KeywordTok{c}\NormalTok{(}\FloatTok{0.4}\NormalTok{, }\FloatTok{0.6}\NormalTok{),}
    \DataTypeTok{recip.prob =} \KeywordTok{matrix}\NormalTok{(}\KeywordTok{c}\NormalTok{(}\FloatTok{0.4}\NormalTok{, }\FloatTok{0.1}\NormalTok{, }\FloatTok{0.2}\NormalTok{, }\FloatTok{0.5}\NormalTok{), }\DataTypeTok{nrow =} \DecValTok{2}\NormalTok{, }\DataTypeTok{byrow =} \OtherTok{TRUE}\NormalTok{))}
\end{Highlighting}
\end{Shaded}

By default, all nodes in the seed network are assumed to be from
group~1. This configuration can be easily customized
as shown in the following example, where nodes~1
and~4 are from group~1, while nodes~2 and~3 are from group~2.

\begin{Shaded}
\begin{Highlighting}[]
\NormalTok{netwk0 \textless{}{-}}\StringTok{ }\KeywordTok{list}\NormalTok{(}\DataTypeTok{edgelist =} \KeywordTok{matrix}\NormalTok{(}\KeywordTok{c}\NormalTok{(}\DecValTok{1}\NormalTok{, }\DecValTok{2}\NormalTok{, }\DecValTok{3}\NormalTok{, }\DecValTok{4}\NormalTok{), }\DataTypeTok{nrow =} \DecValTok{2}\NormalTok{, }\DataTypeTok{byrow =} \OtherTok{TRUE}\NormalTok{),}
    \DataTypeTok{edgeweight =} \KeywordTok{c}\NormalTok{(}\FloatTok{0.5}\NormalTok{, }\DecValTok{2}\NormalTok{), }\DataTypeTok{directed =} \OtherTok{TRUE}\NormalTok{, }\DataTypeTok{nodegroup =} \KeywordTok{c}\NormalTok{(}\DecValTok{1}\NormalTok{, }\DecValTok{2}\NormalTok{, }\DecValTok{2}\NormalTok{, }\DecValTok{1}\NormalTok{))}
\NormalTok{netwk5 \textless{}{-}}\StringTok{ }\KeywordTok{rpanet}\NormalTok{(}\FloatTok{1e3}\NormalTok{, }\DataTypeTok{control =}\NormalTok{ ctr5, }\DataTypeTok{initial.network =}\NormalTok{ netwk0)}
\end{Highlighting}
\end{Shaded}

Node groups are recorded in the data frame \code{node.attr}; immediate
reciprocal edges are denoted as scenario 6 in the data frame \code{edge.attr}.

\paragraph{Customized preference functions}

User-defined preference functions in \proglang{C++} syntax are allowed by
setting \code{ftype = "customized"} in \code{rpa\_control\_preference()}.
This is implemented in \proglang{C++} through the utility functions in
\proglang{R} package \pkg{RcppXPtrUtils}~\citep{Rpkg:RcppXPtrUtils}.
For directed networks,
one-line \proglang{C++} expressions can be passed to arguments \code{spref} and
\code{tpref} to define the source and target preference functions, respectively.
The expressions are strings in \proglang{R} but with valid \proglang{C++} syntax
as transformations of \code{outs} and \code{ins}. The strings are passed to the
function \code{cppXPtr()} in package \pkg{RcppXPtrUtils}, which
compiles the source code and returns an \code{XPtr} (external
pointer) that points to the compiled preference function. The default preference
functions \(f_1(x, y) = x + 1\) and \(f_2(x, y) = y + 1\) can be equivalently
achieved by setting \code{spref = "outs + 1"} and \code{tpref = "ins + 1"}.
The following example sets the preference functions to
\(f_{1}(x, y) = \ln(x + 1) + 1\),
\(f_{2}(x, y) = \ln(y + 1) + 1\):

\begin{Shaded}
\begin{Highlighting}[]
\NormalTok{ctr6 \textless{}{-}}\StringTok{ }\NormalTok{ctr5 }\OperatorTok{+}\StringTok{ }\KeywordTok{rpa\_control\_preference}\NormalTok{(}\DataTypeTok{ftype =} \StringTok{"customized"}\NormalTok{, }
    \DataTypeTok{spref =} \StringTok{"log(outs + 1) + 1"}\NormalTok{, }\DataTypeTok{tpref =} \StringTok{"log(ins + 1) + 1"}\NormalTok{)}
\end{Highlighting}
\end{Shaded}

For undirected networks, argument \code{pref} specifies a one-line
\proglang{C++} expression as a transformation of node strength \code{s}.
The default preference function \(g(x) = x + 1\) could be equivalently achieved by
\code{pref = "s + 1"}.
Users need to ensure the non-negativity of the preference functions.
The generation process will be terminated if a negative preference value is
encountered.

For more advanced preference functions which may take multiple lines of
\proglang{C++} code,
see examples in Appendix~\ref{appendix:customized}.

\section{Benchmarks}
\label{sec:benchmarks}

In this section, we generate weighted and unweighted PA networks with
different sizes and preference functions via our package
(\pkg{wdnet}, version 1.2.0),
\pkg{igraph} (version 1.3.5) and
\pkg{PAFit} (version 1.2.5), and compare their performance.
All simulations were run on a single core of Intel Xeon Gold 6150
CPU @ 2.70GHz with \(16\) GB of RAM.

\paragraph{Weighted networks}

Since the other two packages (i.e., \pkg{igraph} and \pkg{PAFit}) do
not admit edge weights, the comparison of weighted PA network
generation is between the \code{linear} and \code{binary} methods in our
package \pkg{wdnet}. Specifically, we assign the same probabilities
to edge creation scenarios (i.e., \(\alpha = \beta = \gamma = 1/3\)),
set the source and target preference functions respectively to
\(f_{1}(x, y) = x^k + 0.1\) and \(f_{2}(x, y) = y^k + 0.1\)
with \(k \in \left\{0.5, 1, 2\right\}\) (where \(k = 0.5\) and \(k = 2\)
respectively refer to sub-linearity and super-linearity). Draw
the edge weights independently from \({\rm Gamma}(5, 0.2)\). For each
\(k\), we generate PA networks of various evolutionary steps (i.e.,
\(n \in \left\{10^3, \cdots, 10^7\right\}\)) with a simple initial network
consisting of two nodes and one edge \((1, 2, 1)\) (default).

\begin{figure}[tbp]
  \centering
  \includegraphics[width=\linewidth]{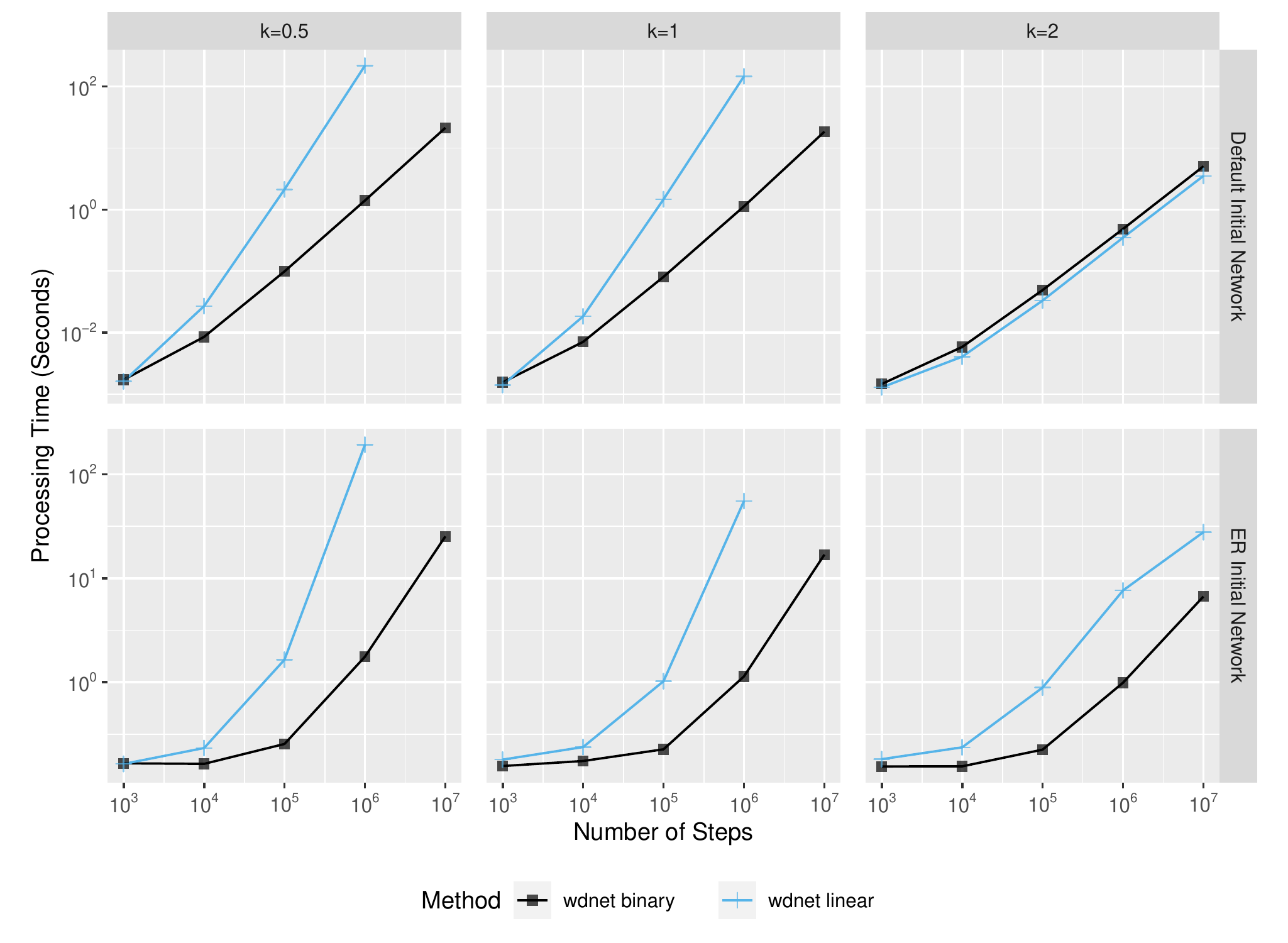}
  \caption{Algorithm runtime for weighted networks
    with default initial network (upper) of one edge $(1, 2, 1)$ and 
    with initial weighted ER networks (lower) of $10^4$ nodes and 
    $10^6$ edges.
    Edge weights, including those in the initial weighted ER networks,
    are drawn from {\rm Gamma}(5, 0.2).
    Probability of edge schemes are $\alpha = \beta = \gamma = 1/3$.
    Preference functions are 
    $f_{1}(x, y) = x^k + 0.1$, $f_{2}(x, y) = y^k + 0.1$ 
    with $k \in \left\{0.5, 1, 2\right\}$.
    Each point represents the median runtime of 100 replications.}
  \label{fig:runtime_weighted_median}
\end{figure}

The top three panels of Figure~\ref{fig:runtime_weighted_median}
compare the median runtimes of generating \(100\) independent weighted,
directed PA networks via \code{binary} and \code{linear} methods.
When preference functions are sub-linear
(\(k = 0.5\)) or linear (\(k = 1\)), the \code{binary} method is
much more efficient than the \code{linear} method. Besides, the
larger the number of steps is, the more advantageous it is to use
the \code{binary} method.
Some simulations for the \code{linear} method are omitted because
they are excessively time-consuming. For a super-linear
preference function (\(k = 2\)), the difference in
generation speed between the two methods becomes subtle. A further investigation
reveals that the sum of source (and target) preference scores of the 20 earliest created
nodes, \(\{v_1, v_2, \ldots, v_{20}\}\), take \(99\%\) of the
total (for all nodes), making them dominant in the sampling process.
Those early created nodes are always quickly selected under whichever edge addition
scenario since linear search visits those ``ancestors'\,' first.
Consequently, the time cost of
using \code{linear} method is significantly reduced.

To further investigate the impact of early created nodes in the sampling process, we
consider a modified (i.e., weighted and directed)
Erd\"{o}s--R\'{e}nyi (ER) network~\citep{Erdos1959on, 
Gilbert1959random} as an initial network. For each
simulation run, we generate an ER network with
\(10^4\) nodes and \(10^6\) edges, where
edge weights are drawn independently from \({\rm Gamma}(5, 0.2)\). We
keep all other parameters same as in the previous experiment,
and give the median runtime (of \(100\) independently generated PA
network replica) in the bottom three panels of
Figure~\ref{fig:runtime_weighted_median}. We observe similar
patterns for \(k = 0.5\) and \(k = 1\), so focus on \(k = 2\) only.
Here the \code{binary} algorithm outperforms the
\code{linear} method, since the large seed network alleviates the
domination of old nodes in the subsequent sampling process.

In fact,
most nodes that are sampled throughout the process are those with
high strengths in the seed graph. Owing to the feature of
ER network, a few hundred nodes (out of \(10^4\)) in the
seed network are repeatedly sampled. However, a larger pool
(compared to \(20\) in the previous experiment) results in longer
runtime when using the \code{linear} method. On the other hand, we
believe the performance of \code{linear} algorithm will improve as
\(n\) gets larger since fewer nodes will continue to be
dominant in the sampling process. Since we have added a sorting
procedure in the \code{linear} algorithm, those nodes will be quickly
selected. Accordingly,
the \code{linear} method may finally outperform the \code{binary}
method for extremely large networks. Last but
not least, we find the tracing curves between \(10^3\) and \(10^4\) become
flatter in the bottom plots when compared to their upper
counterparts (for each \(k\)). This is due to the large
seed graph, which requires a certain amount of time to initialize the
sampling process. Consequently, there is a small difference in the
total generation time for relatively small \(n\).

\paragraph{Unweighted networks}

\begin{figure}[tbp]
  \centering
  \includegraphics[width=\linewidth]{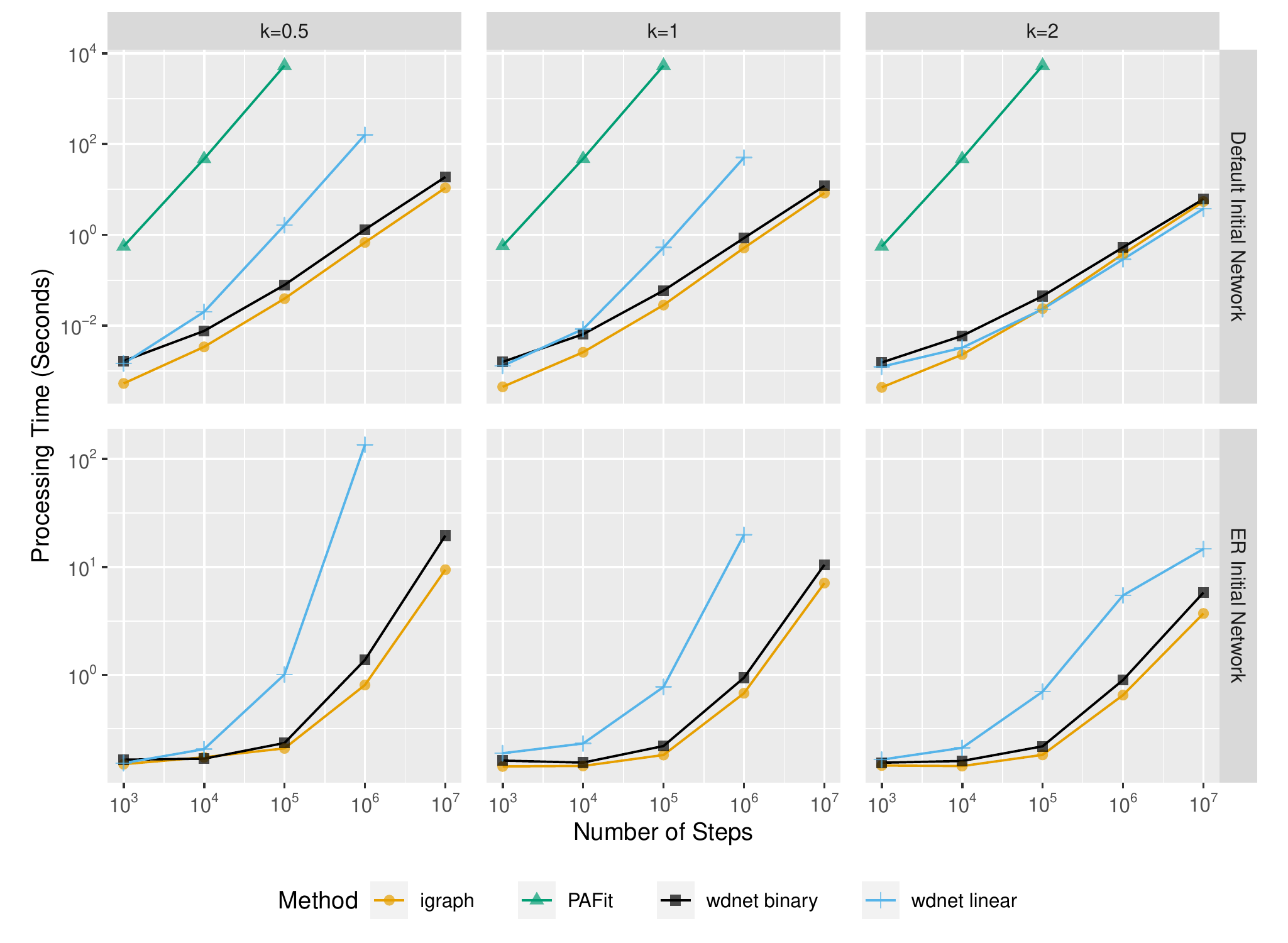}
  \caption{Algorithm runtime for unweighted networks 
    with default initial network (upper) of one edge 
    from $v_1$ to $v_2$ and 
    with initial unweighted ER networks (lower) of $10^4$ nodes and 
    $10^6$ edges.
    Probability of edge schemes are $\alpha = 1$, $\beta = \gamma = 0$.
    The target preference function is 
    $f_{2}(x, y) = y^k + 0.1$.
    Each point represents the median runtime of 100 replications. }
  \label{fig:runtime_unweighted_median}
\end{figure}

Next, we compare the performance of generating unweighted PA
networks using our package \pkg{wdnet} and the other two popular
packages \pkg{PAFit} and \pkg{igraph}. Since \pkg{PAFit} and
\pkg{igraph} allow the \(\alpha\) scheme only, we now
set \(\alpha = 1\) in the \code{rpa\_control\_scenario()} function. Under such
setting, we only need to define a target preference function in the
form of \(f_{2}(x, y) = y^k + 0.1\), where we again assume \(k\in \left\{0.5, 1, 2\right\}\). Similar to the previous
experiments, we generate unweighted PA networks with
\(n \in \left\{10^3, \cdots, 10^7\right\}\) for each \(k\).
A default initial network is adopted for each
simulation run. The results are given in the top three panels of
Figure~\ref{fig:runtime_unweighted_median}. For \(k = 0.5\) and
\(k = 1\), we find the \code{binary} algorithm in our package and
\pkg{igraph} (\code{psumtree} method)
are almost equally efficient, and
outperform the rest. The performance of \code{linear} method in our
package is better than \pkg{PAFit} (similar to linear search) since
the former is implemented in \proglang{C++} whereas the latter is
implemented in \proglang{R}. For \(k = 2\), we do not
see much difference among the two methods in our package \pkg{wdnet}
and that in \pkg{igraph}, which is consistent with our conclusions for
the weighted PA network generation experiment. Overall,
\pkg{PAFit} is the least efficient, especially for generating
large PA networks.

Lastly, we repeat our simulations by considering large ER networks
as the initial graphs in order to alleviate the impact of few old nodes
during the generation process. The setup is the same as that for
weighted network simulations. Noticing that \pkg{PAFit} does not
accept arbitrary initial networks, we exclude it from this set of
simulation comparisons. The corresponding results are shown in the bottom three
panels of Figure~\ref{fig:runtime_unweighted_median}. Similar to the
conclusions drawn for weighted PA networks, the
\code{binary} method outperforms the \code{linear} method
in our package owing to the less influence from old nodes when
\(k = 2\). Moreover, there is little performance difference
between the \code{binary} method and \pkg{igraph} across all
considered \(k\) values.

\section{Discussions}
\label{sec:discussion}

Our \proglang{R} package \pkg{wdnet} provides useful tools to
efficiently generate large-scale PA networks. The package admits a
wide range of PA network specifications such as multiple edge addition scenarios,
weighted and directed edges, and reciprocal edges. Our implementations extend
those discussed in \citet{Wan2017fitting} and \citet{Britton2020directed}, most
of which are not available in other existing packages. A distinctive
feature of the package is that it allow users to define their own
preference functions. Our \code{binary} algorithm is efficient for general
situations. Our \code{linear} algorithm outperforms implementations of the same
algorithm in other packages due to its sorting step.
The core implementation is in \proglang{C++} for fast speed.

Efficient generation of PA networks facilitates investigation of PA networks
properties and goodness-of-fit diagnosis in real applications. A PA network is
controlled by many parameters. When theoretical properties, for example,
transitivity and clustering coefficients, are challenging to derive, their
empirical versions can be easily learned from generating many realizations given
the model parameters. When the initial network size is large relative to the
desired PA network size, its impact may not be ignorable and could be studied
through simulations. In real applications, the goodness-of-fit diagnosis of a PA
network can be done by generating many replicates from the fitted PA model and
comparing the observed network statistics with the empirical distribution of the
same statistics from the replicates. Such goodness-of-fit check may motivate
modification of the PA networks to fit the real data better
\citep[e.g.,][]{Wang2022generating}.

Beyond general PA network generation, \pkg{wdnet} also provides a collection
of other functions. Specifically, several centrality measures are
available via function \code{centrality()}, including the recently
proposed weighted PageRank centrality~\citep{Zhang2022pagerank}.
Assortativity measures for weighted
directed networks discussed in~\citep{Yuan2021assortativity} are
available via function \code{assortcoef()}. A degree-preserving
rewiring algorithm for generating networks with pre-determined assortativity
coefficients~\citep{Wang2022generating} is available via function
\code{dprewire()}.
All these functions are derived from recent research, so they are not available
in other packages.

\section*{Supplementary material}

\begin{enumerate}
  \item[(1)] The code used for benchmarks and the \proglang{R Markdown} source for the paper can be found at 
  \url{https://github.com/Yelie-Yuan/code-sharing/tree/main/generating-pa}.
  \item[(2)] The development version of the package is available at
  \url{https://gitlab.com/wdnetwork/wdnet}.
\end{enumerate}

\section*{Funding}

Dr.~Wang and Dr.~Yan's works were partially supported by the NSF grant
DMS2210735.

\appendix

\section{Alternative sampling method for special cases}
\label{appendix:alternative}

Fast sampling is available when source (target) preference functions
are linear.
We demonstrate this approach through an example of sampling source nodes with
a preference function \(f_{1}(x, y) = x + a_5\).

As shown in Section~\ref{sec:pa}, the main idea of sampling is to make
draws from a bag of node labels, where the number of labels is equal
to the out-degrees. At step \(t + 1\), generate a random
variable \(U \sim {\rm Unif}(0, \sum_{v_j \in V(t)} \theta_1(v_j, t))\), then the source node (for the new edge) is
randomly drawn from the bag if
\(U \leq \sum_{v_j \in V(t)} \text{O}(v_j, t)\).
Otherwise, the source node is uniformly drawn from all existing
nodes (regardless of their out-degrees). The sampling at each step
has complexity \(O(1)\), thus giving complexity \(O(n)\) for the entire network
generation process. The sampling of target nodes can be done in an
analogous manner, and we call this approach \code{bag} in our
package.

This idea can be generalized to weighted networks. At
step \(t + 1\), the source preference score of node \(v_j\) is
\[\sum_{k:(v_j, v_k, w_{jk}) \in E(t)} w_{jk}  + a_5,\]
where total source preference of all existing nodes in the network
is
\[
  \sum_{v_j \in V(t)} \left(\text{O}(v_j, t) + a_5\right) :=
  W(t) + a_5\, |V(t)|,
\]
where \(W(t)\) is the total weight and \(|V(t)|\)
is the cardinality of \(V(t)\). The sampling process proceeds as
follows:

\begin{enumerate}
  \item[(1)] At step $t + 1$, create a vector, $\bm{\nu}(t)$, of 
  cumulative sum of edge weights (according to the emergence order 
  of edges), where the first element is $0$ and the last element is 
  $W(t)$;
  \item[(2)] Compute $\tau(t + 1) = (W(t) + a_5|V(t)|) X$ for some 
  random  variable $X \sim {\rm Unif}(0, 1)$, independent from the
  network generation process;
  \item[(3)] If $\tau(t + 1) > W(t)$, a node is randomly sampled 
  from $V(t)$; otherwise, find an index $\ell$ such that $\tau(t + 
  1) \in (\bm{\nu}_{\ell}(t), \bm{\nu}_{\ell + 1}(t)]$, then select the 
  source node of the edge corresponding to the $\ell$-th addition 
  in $\bm{\nu}(t)$.
\end{enumerate}

The sampling becomes efficient if we apply the above approach to all
steps simultaneously. Notice that edge weights are independently
drawn from \(h\), and they are also independent of other network
generation components. Hence, to efficiently generate networks after \(n\) steps of
evolution, vector \(\bm{\nu}(n)\) can be determined
independently in advance.
Moreover, given a generated list of edge scenario parameters
(i.e., \(\alpha\), \(\beta\) and \(\gamma\)), \(|V(t)|\) can be obtained for
\(1\le t\le n\) as well. Therefore, we collect all information that we
need for node sampling in the entire network generation process,
i.e., \(W(t)\) and \(|V(t)|\) for \(1\le t\le n\), with complexity \(O(n)\).
It remains to find the exact
interval covering \(\tau(t + 1)\) in \(\bm{\nu}(t)\) (a subset of
\(\bm{\nu}(n)\)) for \(\tau(t + 1) \le W(t)\), which can be done efficiently using the
\code{findInterval()} function (with time complexity \(O(n\log{n})\)).

We wrap up the above sampling approach
as \code{bagx} in our package. Although \code{bagx}
is not as efficient as \code{bag} for generating unweighted, linear PA
networks, it provides a competitive alternative to generating
weighted PA networks compared with the standard algorithm.

\section{Advanced customized preference functions}
\label{appendix:customized}

Users can define customized preference functions by
utilizing \code{cppXPtr} from package \pkg{RcppXPrtUtils}. The
returned (external)
pointer, \code{XPtr}, can be passed to \code{spref} and/or
\code{tpref}. For instance, we fix the target preference function as
\(f_{2}(x, y) = y + 1\), and set the source preference function to be
\begin{equation*}
  f_{1}(x, y) = 
  \begin{cases}
    1            & \qquad \text{if } x < 1;\\
    x^2          & \qquad \text{if } 1 \leq x \leq 100;\\
    200 (x - 50) & \qquad \text{otherwise}.
  \end{cases}
\end{equation*}
The corresponding codes are given as follows:

\begin{Shaded}
\begin{Highlighting}[]
\NormalTok{my\_spref \textless{}{-}}\StringTok{ }\NormalTok{RcppXPtrUtils}\OperatorTok{::}\KeywordTok{cppXPtr}\NormalTok{(}\DataTypeTok{code =} 
  \StringTok{"double foo(double x, double y) \{}
\StringTok{    if (x \textless{} 1) \{}
\StringTok{      return 1;}
\StringTok{    \} else if (x \textless{}= 100) \{}
\StringTok{      return pow(x, 2);}
\StringTok{    \} else \{}
\StringTok{      return 200 * (x {-} 50);}
\StringTok{    \}}
\StringTok{  \}"}\NormalTok{)}
\NormalTok{ctr7 \textless{}{-}}\StringTok{ }\KeywordTok{rpa\_control\_preference}\NormalTok{(}\DataTypeTok{ftype =} \StringTok{"customized"}\NormalTok{, }\DataTypeTok{spref =}\NormalTok{ my\_spref, }
  \DataTypeTok{tpref =} \StringTok{"ins + 1"}\NormalTok{)}
\end{Highlighting}
\end{Shaded}

External pointers cannot be shared across different \proglang{R} sessions.
Therefore, we recommend that users save the source code of the customized
preference functions for recompilation.

\bibliography{generatePA.bib}
\bibliographystyle{jds}

\end{document}